# On Landauer vs. Boltzmann and Full Band vs. Effective Mass Evaluation of Thermoelectric Transport Coefficients

Changwook Jeong, Raseong Kim, Mathieu Luisier, Supriyo Datta, and Mark Lundstrom

Network for Computational Nanotechnology

Birck Nanotechnology Center

Purdue University

West Lafayette, Indiana, 47907

**Abstract-** The Landauer approach to diffusive transport is mathematically related to the solution of the Boltzmann transport equation, and expressions for the thermoelectric parameters in both formalisms are presented. Quantum mechanical and semiclassical techniques to obtain from a full description of the bandstructure, $E(k)$, the number of conducting channels in the Landauer approach or the transport distribution in the Boltzmann solution are developed and compared. Thermoelectric transport coefficients are evaluated from an atomistic level, full band description of a crystal. Several example calculations for representative bulk materials are presented, and the full band results are related to the more common effective mass formalism. Finally, given a full $E(k)$ for a crystal, a procedure to extract an accurate, effective mass level description is presented.

**1) Introduction**

Much experimental and theoretical effort has been directed at improving the thermoelectric (TE) figure of merit, $ZT = S^2 GT/K$, where $T$ is the temperature, $S$ is the Seebeck coefficient, $G$ is the electrical conductance, and $K$ is the thermal conductance, which is the sum of the electronic contribution, $K_e$, and the lattice thermal conductance, $K_l$. Careful tradeoffs are needed to obtain high ZT. Recent experimental reports of high ZT[1-8] are attributed to suppressing the lattice thermal conductivity, and now the question of whether the electronic performance can be enhanced is being asked[9-11]. New materials[1,12-15], new structures (e.g. nanowires[16-26], quantum wells[2,27,28], superlattices[10,11,17,29-36], and nanocomposites[3,4,37-39]), and strain engineering[29,40-42], which has been so successful for enhancing the electronic performance of nanotransistors, are all being explored. To address these opportunities, thermoelectric coefficients must be related to an accurate description of the electronic structure of the material.

Thermoelectric parameters are usually evaluated by solving the Boltzmann Transport Equation (BTE)[43]. For low temperature thermoelectrics in mesoscopic structures, the Landauer approach is commonly used[44,45]. The Landauer approach applies to high temperature diffusive samples as well, and it provides an alternative formulation that can be insightful[46]. One objective for this paper is to discuss the mathematical relation between the Landauer and Boltzmann approaches.

In both the Landauer and Boltzmann approaches the thermoelectric parameters are related to the electronic structure of the material. The effective mass approach is widely-used to analyze experiments and to design devices. For more complex materials, full band treatments (*ab initio* or empirical tight binding) have been used [12,47-56]. It is still not clear, however, exactly how full band treatments relate to effective mass level treatments

– especially for complex bandstructures. Another objective of this paper is to discuss the evaluation of thermoelectric parameters from a full band perspective and to show that the results are easily related to an effective mass level description.

The paper is organized as follows. In Sec. 2, we present a brief summary of the Landauer formalism and relate it to the more common approach that begins with the BTE. We also present two methods for evaluating the transport distribution in the Landauer approach from a full band description of the electronic structure. In Sec. 3, tight binding simulation results are presented for the conduction and valence bands of germanium (Ge), gallium arsenide (GaAs), and bismuth telluride ($Bi_2Te_3$). The results are discussed within the Landauer framework in Sec. 4, as is the relation of the rigorous approach to the effective mass approach. Our conclusions are summarized in Sec. 5.

**2) Approach**

The Landauer formalism in the linear response regime gives the electrical conductance, Seebeck coefficient, and the thermal conductance for zero electric current as

$$G = \left(2q^2/h\right)I_0 \qquad [1/\Omega] \qquad (1)$$

$$S = \left[k_B/(-q)\right]\frac{I_1}{I_0} \qquad [V/K] \qquad (2)$$

$$K_e = \left(T_L 2k_B^2/h\right)\left\{I_2 - I_1^2/I_0\right\}, \qquad [W/K] \qquad (3)$$

where

$$I_j = \int_{-\infty}^{+\infty} \left(\frac{E - E_F}{k_B T_L}\right)^j \bar{T}(E)\left(-\frac{\partial f_0}{\partial E}\right) dE, \qquad (4)$$

with

$$\bar{T}(E) = T(E)M(E),  \quad (5)$$

being the transmission[45], and $M(E)$ the number of conducting channels. For a conductor of length, $L$, and mean-free-path for backscattering, $\langle\langle\lambda(E)\rangle\rangle$,

$$T(E) = \langle\langle\lambda(E)\rangle\rangle/L  \quad (6)$$

in the diffusive limit[57]. For some common scattering mechanisms, $\langle\langle\lambda(E)\rangle\rangle$ can be expressed in power law form as $\langle\langle\lambda(E)\rangle\rangle = \lambda_0 (E/k_B T)^r$, where $\lambda_0$ is a constant, $E$ is the kinetic energy, and $r$ is a characteristic exponent describing a specific scattering process.

Thermoelectric transport coefficients are more commonly obtained by solving the Boltzmann equation in the relaxation time approximation and expressed in terms of an integral like eqn. (4) with the transmission replaced by the so-called transport distribution according to

$$\Sigma(E) = \frac{L^2}{h} M(E) T(E) = \frac{L}{h} M(E) \langle\langle\lambda\rangle\rangle  \quad (7)$$

A solution of the Boltzmann equation gives [43]

$$\Sigma(E) = \sum_{\vec{k}} \left(v_x^2 \tau\right) \delta(E - E_k)  \quad (8)$$

where $\tau$ is the relaxation time. Equation (7) relates the solution of the Boltzmann equation in the relaxation time approximation to the Landauer formalism.

By making the definition

$$\langle |v_x| \rangle \equiv \frac{\sum_{\vec{k}} |v_x| \delta(E - E_k)}{\sum_{\vec{k}} \delta(E - E_k)} \quad (9)$$

eqn. (8) can be expressed as

$$\Sigma(E) = \frac{\langle v_x^2 \tau \rangle}{\langle |v_x| \rangle} \langle |v_x| \rangle D(E) = \frac{\langle v_x^2 \tau \rangle}{\langle |v_x| \rangle} \sum_k |v_x| \delta(E - E_k) \quad (10)$$

where $D(E) = \sum_k \delta(E - E_k)$ is the density of states.

Finally, according to eqn. (7), we find[57]

$$M(E) = \frac{h}{2L} \sum_k |v_x| \delta(E - E_k) \quad (11)$$

and

$$\langle\langle \lambda(E) \rangle\rangle = 2 \frac{\langle v_x^2 \tau \rangle}{\langle |v_x| \rangle} \quad (12)$$

Equation (11) relates the number of conducting channels in the Landauer formalism[45] to the bandstructure. Equation (12) is an appropriately defined mean-free-path (the mean-free-path for backscattering) so that the Landauer results agree with the Boltzmann equation in the relaxation time approximation. Assuming isotropic energy bands, eqn. (12) can be evaluated in one-dimension (1D), two-dimensions (2D), and three-dimensions (3D) to find

$$\langle\langle \lambda(E) \rangle\rangle = 2v(E)\tau(E) \quad (1D) \quad (13a)$$

$$\langle\langle \lambda(E) \rangle\rangle = (\pi/2)v(E)\tau(E) \quad (2D) \quad (13b)$$

$$\langle\langle\lambda(E)\rangle\rangle = (4/3)\upsilon(E)\tau(E) \quad (3D) \tag{13c}$$

In practice, a constant scattering time is often assumed for the Boltzmann equation, but this is hard to justify. In the Landauer approach, a constant mean-free-path simplifies calculations and can be justified in 3D for parabolic bands when the scattering rate is proportional to the density of states.

The discussion above shows that $M(E)$ is essentially the carrier velocity times the density-of-states. If we consider a single parabolic conduction band, $E = \hbar^2 k^2 / 2m^*$, then $M(E)$ for 3D is

$$M(E) = A \frac{m^*_{DOM}}{2\pi\hbar^2} E \tag{14}$$

where the density-of-modes effective mass is just $m^*$ for a single, spherical band. (Results for 1D and 2D are given in Ref. 46) For ellipsoidal energy bands, eqn. (11) can be evaluated for each equivalent ellipsoid to find $m^*_{DOM} = \sqrt{m^*_y m^*_z}$ with the direction of current flow being along the x-direction. This example shows that $M(E)$ is related to the density-of-states in the 2D plane transverse to the transport direction. The contributions for each equivalent ellipsoid are then summed. For the conduction band of silicon, the result is $m^*_{DOM} = 2m^*_t + 4\sqrt{m^*_t m^*_l}$ which is 2.04 $m_0$. Recall that the density of states effective mass is $m^*_{DOS} = 6^{2/3}(m_l m_t^2)^{1/3} = 1.06 m_0$. This example shows that the density-of-modes and density-of-state effective masses can be quite different. Finally, for non-parabolic bands with Kane's dispersion relation[58], $E(1+\alpha E) = \hbar^2 k^2/2m^*$, $M(E)$ for 3D becomes

$$M(E) = A \frac{m^*}{2\pi \hbar^2} E(1+\alpha E), \qquad (15)$$

where $\alpha$ is the non-parabolicity parameter. These analytical results will be our reference against which we compare the numerical results to be presented later.

Two procedures are available to numerically evaluate $M(E)$. Firstly, $M(E)$ can be calculated by counting bands for a given bandstructure, because we can express eqn. (11) as [57,59]

$$M(E) = \sum_{k_\perp} \Theta(E - E_{k_\perp}), \qquad (16)$$

where $\Theta$ is the unit step function and $k_\perp$ refers to $k$ states perpendicular to the transport direction (i.e., transverse modes). Equation (16) is simply a count of the bands that cross the energy of interest and provides a computationally simple way to obtain $M(E)$ from a given $E(k)$. Similar expressions have been used to numerically evaluate the number of modes for phonon transport from a given dispersion relations[60]. A MATLAB® script that implements this calculation for Ge is available[61].

An alternative to counting the number of available bands at a given energy consists of calculating the transmission coefficient through a given structure as function of the injection energy. In the non-equilibrium Green's function formalism[57], $\bar{T}(E)$ is

$$\bar{T}(E) = Tr(\Gamma_1 G \Gamma_2 G^\dagger), \qquad (17)$$

where *G* is the retarded Green's function and

$$\Gamma_{1,2} = i(\Sigma_{1,2} - \Sigma_{1,2}^{\dagger}) \tag{18}$$

where $\Sigma_{1,2}$ are the contact self-energies. This approach works for bulk thermoelectrics, but it also allows us to obtain the TE parameters for quantum-engineered structures for which the electronic structure may be very different from the bulk.

For our calculations, we have developed a multi-dimensional quantum transport simulator based on different flavors of the nearest-neighbor tight-binding model. It solves Schrödinger equation in the Wave Function (WF) formalism, which in the ballistic limit is equivalent to the Non-equilibrium Green's Function (NEGF), but computationally much more efficient[62]. To obtain the bulk transmission coefficient $\bar{T}(E)$, a small device structure composed of two to three unit cells is constructed, two semi-infinite contacts are attached to both ends of the simulation domain, and electrons and holes are injected and collected from these contacts. This procedure is repeated for different energies and wave vectors so that the entire Brillouin Zone of the considered semiconductor material is spanned. We integrate the resulting transmission coefficient over its momentum-dependence at a given energy to evaluate $\bar{T}(E)$.

To evaluate *M(E)* beyond the effective mass approximation, an accurate description of the electronic structure is needed. Materials like Si, Ge, or GaAs have been parametrized in the nearest-neighbor tight-binding (TB) model by several groups[63-65] with different levels of approximation (e.g. $sp^3s^{*}$ [66] and $sp^3d^5s^{*}$ [65] models) for many years. More exotic materials like $Bi_2Te_3$ been parametrized[47,48]. A comparison with energy bands obtained from Density Functional Theory (DFT) shows that a nearest-neighbor $sp^3d^5s^*$ tight-binding approach with spin-orbit coupling is required to capture the essential characteristics of the $Bi_2Te_3$ bandstructure[48]. Hence, we have extended our quantum transport simulator described above to include the rhombohedral crystal lattice and to calculate transmission coefficients through such structures.

**3) Results**

In this section, we illustrate the techniques discussed in Sec. 2 and show how full band calculations are related to effective mass calculations. A few materials that are good illustrations (not necessarily good TE materials) are compared: a) Ge to compare 3 approaches to compute the number of modes - counting bands, NEGF-TB model, effective mass approximation (EMA) – which should all agree rather well since the Ge conduction bands are nearly parabolic, b) Ge valence band to see if we can use an effective mass description for the valence band, c) GaAs to illustrate the effect of non-parabolicity, and d) $Bi_2Te_3$ because it is commonly used thermoelectric with a more complex bandstructure.

Figure 1 shows the number of modes, $M(E)$ for the Ge conduction band as computed by 3 different approaches. Counting bands gives exactly the same $M(E)$ obtained by NEGF-TB model. As shown in Fig. 1, the EMA expression for $M(E)$ (eqn. (14)) provides a good fit to the full band calculation near the conduction band edge. Full band calculations of the density of states, $D(E)$ and $M(E)$ for Ge, GaAs, and $Bi_2Te_3$ are shown in Fig. 2. Around the band edge, the linear density of modes ($M(E)$) vs. energy expected from eqn. (14) is observed for all materials considered – even for the highly warped valence band. In the bulk, $M(E)$ varies linearly with $E$ because both $D(E)$ and $\upsilon(E)$ are proportional to $\sqrt{E}$. A linear behavior of the "transport distribution" $\Sigma(E)$ vs. $E$ has previously been observed[49], but the transport distribution varies as $D(E)$ times $\upsilon^2(E)$, so it is not expected to be exactly linear when the relaxation time, $\tau$, is assumed to be constant.

To show the relation between full band calculation and the EMA, a "density-of-modes" effective mass ($m^*_{DOM}$) was extracted from the numerically evaluated $M(E)$ using eqn. (14) and compared to the analytical $m^*_{DOM}$ with number of valleys and transport direction being accounted for. The results are listed in Table 1. The discrepancy is no larger than 10% for conduction band, while it is about a factor of 2 for valence band. As shown in Fig. 3 for the conduction band of GaAs, a better fit can be obtained when non-parabolicity is accounted for, and the discrepancy between extracted $m^*_{DOM}$ and analytic one reduced from 10% to 2%. As listed in Table 1 and discussed in Sec. 2, the "density-of-states" effective masses are clearly different from the density-of-modes effective masses - except for GaAs, where the Gamma valley is the conduction band minimum. Finally, note that although there is no simple relation between the light and heavy hole effective masses and the numerically extracted $m^*_{DOM}$ for the valence band, a constant $m^*_{DOM}$ provides a good fit to $\bar{T}(E)$.

**4) Discussion**

In this section, thermoelectric properties will be evaluated and interpreted within the Landauer framework. Figure 4 compares calculated Seebeck coefficients (*S*) using eqn. (2) to experiments. The results are plotted vs. reduced Fermi level ($\eta_F = (E_F - E_C)/k_B T$), and we assume that the scattering rate ($1/\tau$) is proportional to the density-of-states, i.e. phonon scattering is dominant[67], which is equivalent to a constant mean-free-path, $\lambda_0$. The Seebeck coefficient (eqn. (2)) is independent of $\lambda_0$. The results clearly demonstrate that *S* is nearly independent of electronic band structure (i.e., of $m^*_{DOM}$). In the effective mass approximation, the Seebeck coefficient in 3D is $S_{3D} = -(k_B/q)\left((r+2)\mathscr{F}_{r+1}(\eta_F)/\mathscr{F}_r(\eta_F) - \eta_F\right)$, which depends only on the location of the Fermi level and on *r*, where *r* is the characteristic exponent that describes scattering. The Seebeck coefficient depends weakly on electronic structure but more strongly on scattering. Ioffe, for example, pointed out the possibility of making use of ionized impurity scattering (*r* = 2) to improve *S* [68].

The constant mean-free-path was adjusted to give the best match to experimental data for electrical conductivity ($\sigma$) with its corresponding *S*. This approach is essentially the same as the common approach in which the unknown relaxation time, $\tau$, is treated as a constant [40,49,48,69], which actually turned out to be good approximation even for systems with crystal anisotropy[49,69]. With the best fit $\lambda_0$, the power factor ($PF = S^2 G$) and electronic thermal conductivity ($\kappa_e$) were then evaluated using eqn. (1-3). The thermoelectric figure of merit, *ZT* was computed at 300 K using calculated values of *PF* and $\kappa_e$ and the experimentally determined the lattice thermal conductivity, $\kappa_l$ [67]. Figure 5 shows well-fitted results for Bi$_2$Te$_3$ with $\lambda_0 = 18, 4$ nm for conduction and valence band, respectively. Figure 6 compares the calculated *PF* and *ZT* vs. Fermi level to experiments for Ge, GaAs and Bi$_2$Te$_3$. Calculated results agree well with experiments. (The parameters used in these calculations are summarized in Table 2.) These results show that the Landauer approach gives essentially the same accuracy as the BTE approach (although the use of a constant mean-free-path is easier to justify than the use of a constant relaxation time). The Landauer approach has the benefit of being readily extendable to ballistic (e.g. thermionic) and to quantum-engineered structures.

We now consider the effective mass level treatment of this problem. To calculate TE coefficients and analyze measured TE data within the EMA, two effective masses are needed: 1) $m^*_{DOM}$ for $M(E)$ calculation 2) $m^*_{DOS}$ to obtain the reduced Fermi-level ($\eta_F = (E_F - E_C)/k_B T$) from measured carrier concentration. In the EMA,

$$S_{3D} = -\frac{k_B}{q} \left( \frac{(r+2)\mathscr{F}_{r+1}(\eta_F)}{\mathscr{F}_r(\eta_F)} - \eta_F \right) \tag{19}$$

$$G_{3D} = \lambda_0 \frac{2q^2}{h} \frac{m^*_{DOM} k_B T}{2\pi \hbar^2} \Gamma(r+2) \mathscr{F}_r(\eta_F) \tag{20}$$

$$K_{e,3D} = \lambda_0 T \left(\frac{k_B}{q}\right)^2 \frac{2q^2}{h} \frac{m^*_{DOM} k_B T}{2\pi \hbar^2} \Gamma(r+3) \left( (r+3)\mathscr{F}_{r+2}(\eta_F) - \frac{(r+2)\mathscr{F}^2_{r+1}(\eta_F)}{\mathscr{F}_r(\eta_F)} \right) \tag{21}$$

where *r* is the characteristic exponent describing a specific scattering mechanism, and $\lambda_0$ is determined by comparison with experiments. Figures 5 and 7 show that effective mass theory provides a good agreement with full band atomistic simulation results.

Because the valence bands are coupled and warped, it is difficult to predict $m^*_{DOM}$ from the values of the heavy- and light- hole effective masses. Indeed, Table 1 shows a large discrepancy between the expected and numerically extracted values. From the Luttinger-Kohn model, the valence band near the $\Gamma$ point can be expressed as[70],

$$E(k) - E_V = \frac{\hbar^2}{2m_x}\left[\gamma_1 k^2 \pm \sqrt{4\gamma_2^2 k^4 + 12(\gamma_3^2 - \gamma_2^2)(k_x^2 k_y^2 + k_y^2 k_z^2 + k_x^2 k_z^2)}\right] \quad (22)$$
$$= Ak^2 \pm \sqrt{B^2 k^4 + C^2(k_x^2 k_y^2 + k_y^2 k_z^2 + k_x^2 k_z^2)}$$

where $\gamma_i$ are the Luttinger parameters and A, B, and C are constants. From the definition of number of modes, eqn. (11), it is hard to derive analytically the $M(E)$ vs. E relation and then find analytical expression for $m^*_{DOM}$. But based on the counting bands approach, we can readily see why the extracted $m^*_{DOM}$ is about two times larger than expected one from EMA.

Figure 8(a) shows that the conduction band of GaAs is nearly parabolic near the $\Gamma$ point. According to the counting bands approach, e.g. eqn. (16), each band gives one conducting mode for electrons at a specific energy, *E*, due to parabolic behavior of dispersion relation. In other words, effective mass approximation assumes that each band gives one conducting channel for an injected electron having a specific wave vectors and energy *E*. When the bands are nearly parabolic, the analytic $m^*_{DOM}$ agrees well with the $m^*_{DOM}$ extracted from full band calculation, as we can see for the conduction band in Table 1.

If we assume parabolic bands for the valence band (heavy- and light hole) close to the Γ point, the $m^*_{DOM}$ is expressed as $m^*_{DOM} = m_{lh} + m_{hh}$, which is approximately two times less than the value extracted from full band calculation as shown in Table 1. As clearly shown in the Fig. 8(b), most of bands for holes (especially for heavy-hole) contribute at least two conducting channels at a specific energy. The parabolic band assumption, however, gives one conducting channel per band and significantly underestimates the number of modes for holes. Warped valence bands provide more conducting modes.

Using this argument, we may also explain qualitatively the question of why $m^*_{DOM}$ is different between Ge and GaAs even though the valence bands look similar. Including results for Si and InAs valence bands, the hole 'density-of-modes' effective mass, $m^*_{DOM}$ extracted from full band calculations for Si, Ge, GaAs, and InAs are given as, $2.40m_0 > 1.63m_0 > 0.97m_0 > 0.65m_0$, respectively. In eqn. (22), the degree of warping can be judged from the values of $(\gamma_3^2 - \gamma_2^2)/\gamma_2^2$, which we call the warping parameter. For $\gamma_3 = \gamma_2$, eqn. (22) yields two parabolic bands (heavy- and light – hole). From the tabulated values of $\gamma_i$ [70], the calculated warping parameter are $17 > 0.76 > 0.62 > 0.20$ for Si, Ge, GaAs, and InAs, respectively. This shows that the degree of warping can qualitatively explain the relative magnitude of $m^*_{DOM}$ for Si, Ge, GaAs, and InAs even though the valence band for all those diamond-like materials looks similar. One thing to note is that 6-valley valence band structure of $Bi_2Te_3$ is another reason for its high $m^*_{DOM}$.

**5) Summary and Conclusion**

The relation between the so-called transport distribution, which determines the TE coefficients and begins with the BTE, and the transmission obtained from the Landauer approach has been clarified in this paper. We also showed that the transmission (transport distribution) is readily obtained from the full band description of the electronic bandstructure of a semiconductor using well-developed techniques - a simple semiclassical band counting method and a quantum mechanical approach. Several example calculations of the transmission and the TE coefficients for representative bulk materials were presented to demonstrate that Landauer approach provides an accurate description of experimentally measured thermoelectric parameters, In practice, the use of a constant mean-free-path in the Landauer approach is easier to justify than the use of a constant relaxation time in the Boltzmann equation. The Landauer approach also provides complementary insight into thermoelectric physics and can be applied to ballistic, quasi-ballistic, and quantum engineered structures. Finally, we showed that an accurate and simple effective mass model can be defined by extracting a "density-of-modes" effective mass from the given full band results. One first computes $M(E)$ from eqn. (16) and then fits the linear portion near the band edge to eqn. (14). For accurate results, the fitting should be performed from the band edge to $\approx 5k_BT$ above the maximum expected Fermi level at the highest temperature of operation.

**Acknowledgments**

This work was supported by the Network for Computational Nanotechnology under National Science Foundation Grant No. EEC-0634750.

**Figure Captions**

Figure 1. (a) Comparison of the number of modes, $M(E)$, computed by 3 different approaches for Germanium (Ge): NEGF-TB model, Effective Mass Approximation (EMA), and counting bands. The $M(E)$ from counting bands (dashed line) is on top of $M(E)$ from the NEFG-TB model. (b) Illustration of bands counting method for specific dispersion relation for Ge. Dotted line is guide to eye.

Figure 2. Full band calculations of the density of states (*DOS*) and the number of modes (*M*) for Ge, GaAs, and $Bi_2Te_3$. The midgap is located at E = 0. The inset in Fig. 2(b) shows $M(E)$ near the conduction band edge for GaAs.

Figure 3. Comparison of fitting based on parabolic dispersion relation with fitting based on Kane dispersion relation. Non-parabolicity parameter $\alpha$ used for GaAs is 0.64[71]. Above 1eV, *L* valleys contribute to the number of modes in addition to $\Gamma$ valley.

Figure 4. Calculated Seebeck coefficients (S) using eqn. (2) and experiments[72-74,67] as a function of reduced Fermi level ($\eta_F = (E_F - E_C)/k_BT$). We assumed that scattering rate ($1/\tau$) is proportional to the density-of-states, i.e. phonon scattering is dominant[67]. The reduction of Seebeck coefficient around $\eta_F = -2$ for $Bi_2Te_3$ is attributed to the bipolar conduction due to its relatively small bandgap (0.162 eV).

Figure 5. Comparison of the simulated and experimentally[67] measured *S, G,* and κ for $Bi_2Te_3$ assuming a constant mean-free-path, $\lambda_0 = 18, 4$ nm for conduction and valence bands. Thermal conductivity is the sum of the electronic and lattice thermal conductivity. Used parameters are listed in Table 2.

Figure 6. Calculated and measured PF and ZT as function of the Fermi level. Used parameters are listed in Table 2.

Figure 7. Comparison of EMA with full-band calculation for Ge. On the *y* axis, Seebeck coefficient (*S*), electrical conductivity (*G*) and thermal conductivity by electron($\kappa_e$) are plotted from 0 to 400 μV/K, 0 to 4E6 $\Omega^{-1}m^{-1}$, and 0 to 40 W $m^{-1}$ $K^{-1}$.

Figure 8. Energy dispersion relation showing the lowest (a) conduction bands and (b) valence bands of GaAs. ( y axis ranges from $E_C$ (or $E_V$) to $E_C$ (or $E_V$) + $5k_BT$ because $-\partial f_0/\partial E$ spread about $5k_BT$.) Each red dot represents a conducting channels for positive moving electrons at specific energy for an electron moving with a positive velocity. In the valence bands, most of the bands (especially heavy holes) have at least two conducting channels per energy

**Table Captions**

Table 1. Analytic "density-of-modes" and full band NEGF-TB simulation. For comparison, "density-of-states" effective masses ($m^*_{DOS}$) are also listed. The transport direction is along the *x* direction. The electron ($m^e$) and hole effective masses ($m^{lh}$, $m^{hh}$) in the device coordinate (*x*, *y*, *z*) are used for analytic effective mass calculations and are given in units of the free electron mass. The 'heavy-hole' effective masses (Ge: 0.35 and GaAs: 0.51) assume spherical symmetry[75,76]. The extracted 'heavy-hole' effective mass for Ge and GaAs has a strong anisotropy (Ge: 0.17 [100], 0.37 [110], 0.53 [111], and GaAs: 0.38 [100], 0.66 [110], 0.84 [111] ). CB denotes conduction band and VB denotes valence band. The top three shaded rows are for the conduction bands and generally show good agreement between analytic and numerically extracted values. The bottom three rows for the valence band (VB) generally show a much larger discrepancy. The two columns at the right (enclosed in dashed lines) show that analytic and numerically extracted density-of-states effective masses generally agree reasonably well, but the density-of-states effective masses are typically much lower than the density of modes effective masses.

Table 2. summary of parameters used in Figure 6: fitted $\lambda_0(nm)$ parameters, experimental lattice thermal conductivity $\kappa_l (Wm^{-1}K^{-1})$. In the power law form of the mean free path, $\langle\langle \lambda(E) \rangle\rangle = \lambda_0 \left(E/k_B T\right)^r$, *r* is 0 since we assumed that phonon scattering is dominant.

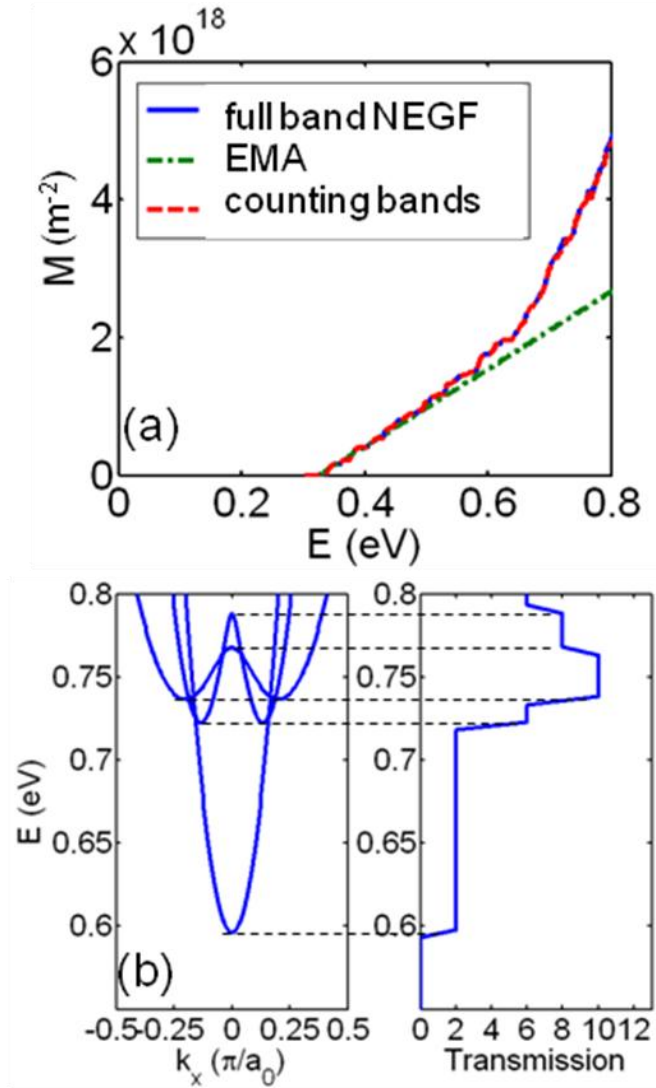

Figure 1. (a) Comparison of the number of modes, $M(E)$, computed by 3 different approaches for Germanium (Ge): NEGF-TB model, Effective Mass Approximation (EMA), and counting bands. The *M(E)* from counting bands (dashed line) is on top of *M(E)* from the NEFG-TB model. (b) Illustration of bands counting method for specific dispersion relation for Ge. Dotted line is guide to eye.

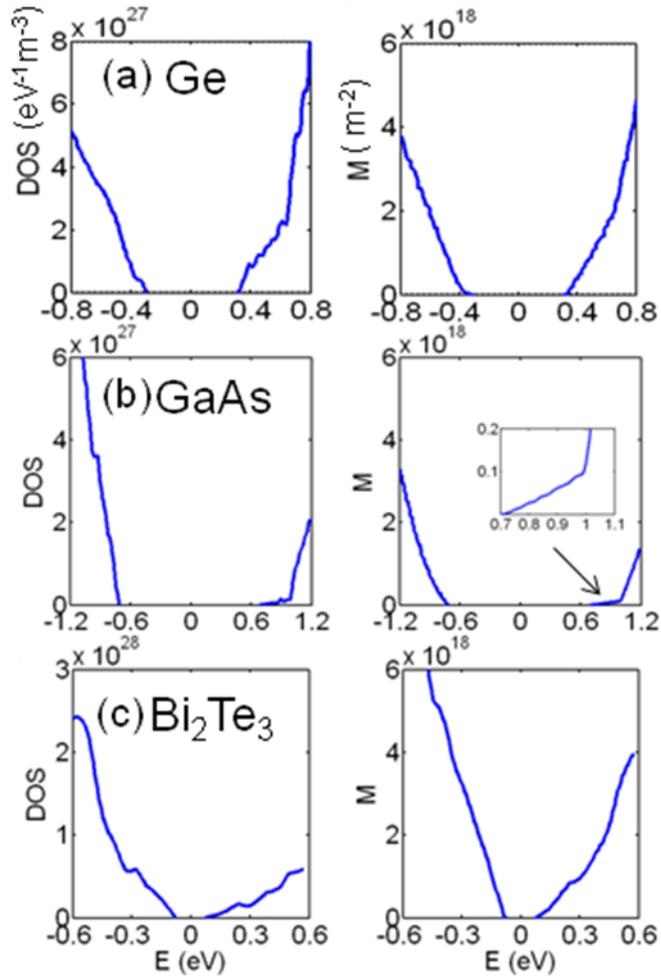

Figure 2. Full band calculations of the density of states (*DOS*) and the number of modes (*M*) for Ge, GaAs, and $Bi_2Te_3$. The midgap is located at E = 0. The inset in Fig. 2(b) shows *M(E)* near the conduction band edge for GaAs.

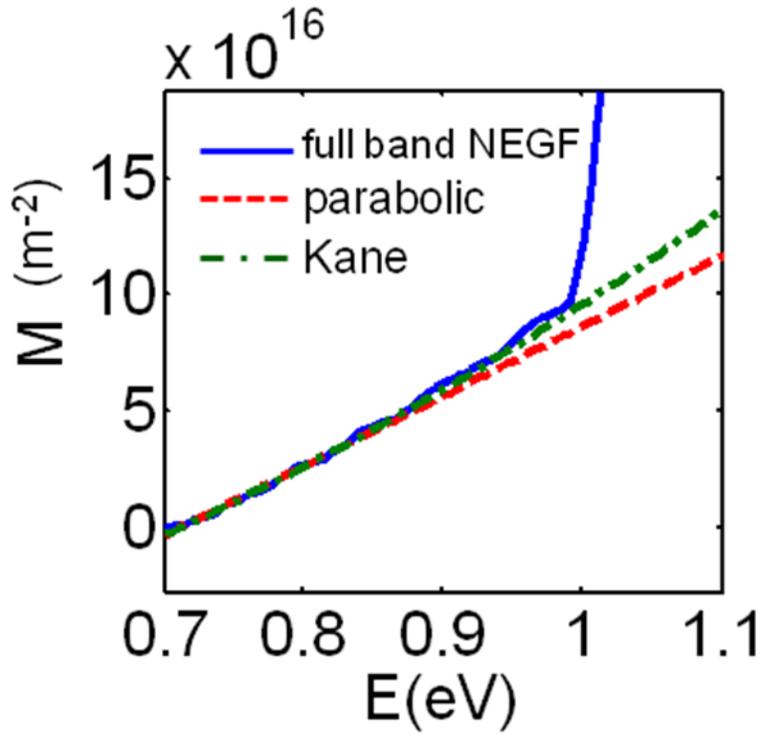

Figure 3. Comparison of fitting based on parabolic dispersion relation with fitting based on Kane dispersion relation. Non-parabolicity parameter $\alpha$ used for GaAs is 0.64[71]. Above 1eV, $L$ valleys contribute to the number of modes in addition to $\Gamma$ valley.

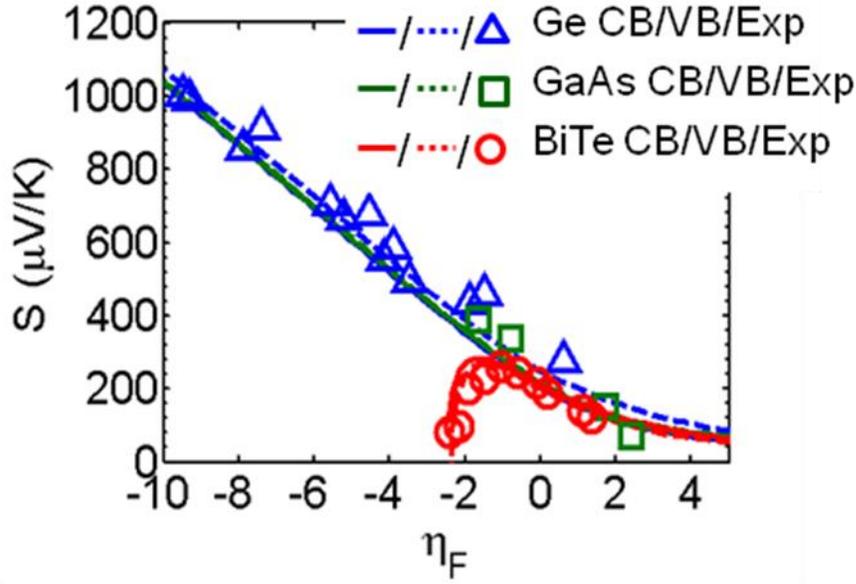

Figure 4. Calculated Seebeck coefficients (S) using eqn. (2) and experiments[72-74,67] as a function of reduced Fermi level ($\eta_F = (E_F - E_C)/k_B T$). We assumed that scattering rate ($1/\tau$) is proportional to the density-of-states, i.e. phonon scattering is dominant[67]. The reduction of Seebeck coefficient around $\eta_F = -2$ for $Bi_2Te_3$ is attributed to the bipolar conduction due to its relatively small bandgap (0.162 eV).

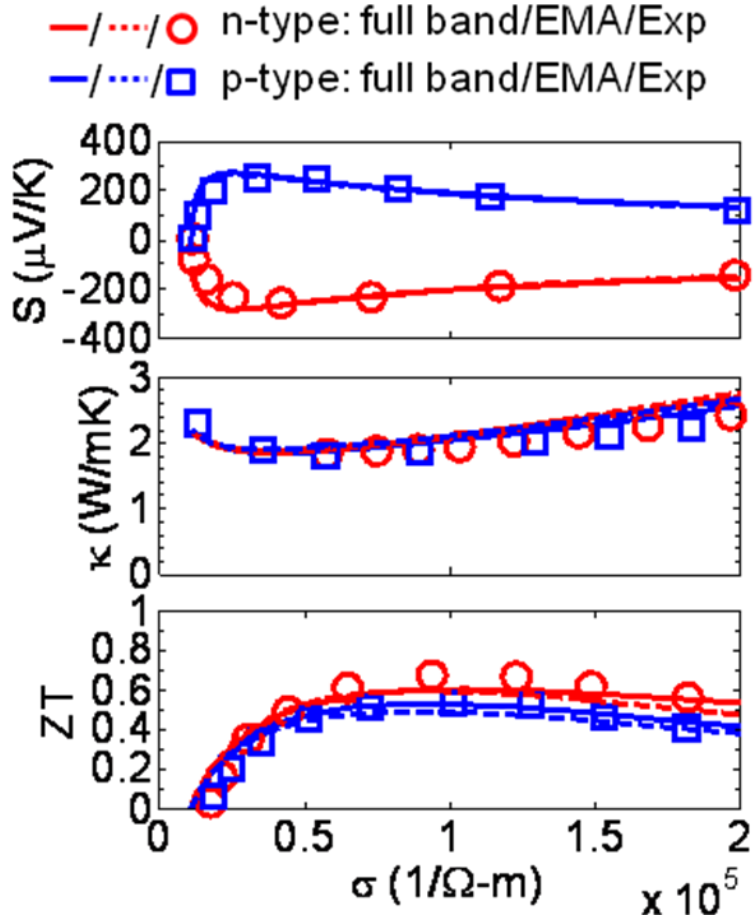

Figure 5. Comparison of the simulated and experimentally[67] measured $S$, $G$, and $\kappa$ for $Bi_2Te_3$ assuming a constant mean-free-path, $\lambda_0 = 18, 4$ nm for conduction and valence bands. Thermal conductivity is the sum of the electronic and lattice thermal conductivity. Used parameters are listed in Table 2.

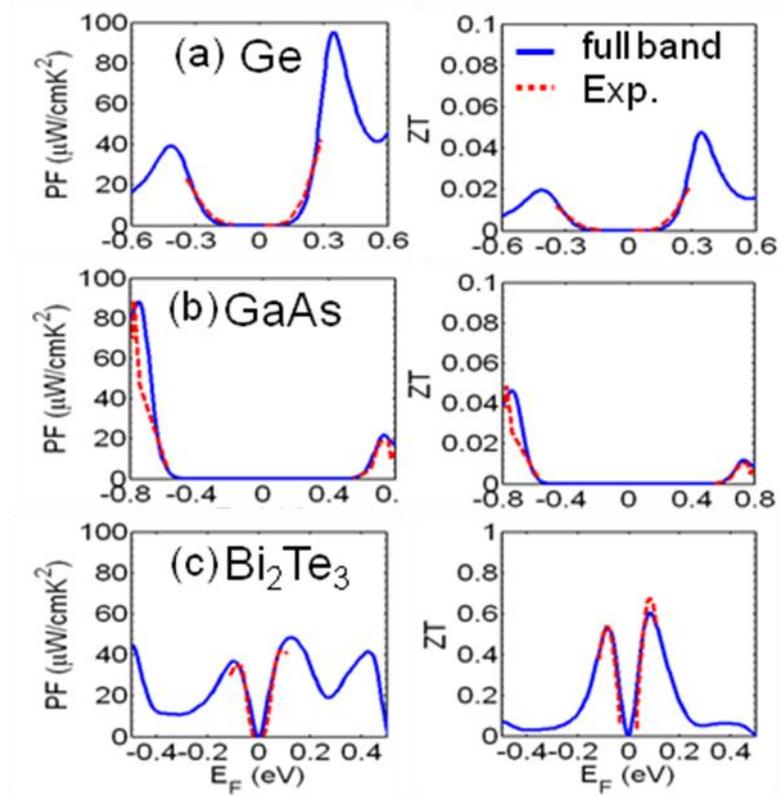

Figure 6. Calculated and measured PF and ZT as function of the Fermi level. Used parameters are listed in Table 2.

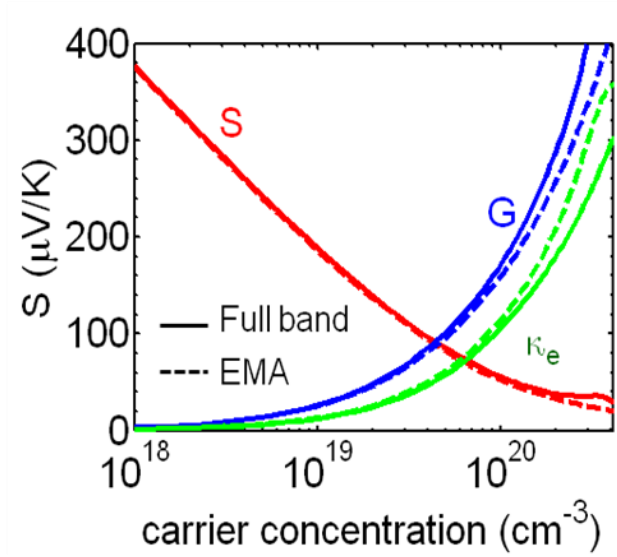

Figure 7. Comparison of EMA with full-band calculation for Ge. On the *y* axis, Seebeck coefficient (*S*), electrical conductivity (*G*) and thermal conductivity by electron ($\kappa_e$) are plotted from 0 to 400 μV/K, 0 to 4E6 $\Omega^{-1}m^{-1}$, and 0 to 40 W m$^{-1}$ K$^{-1}$.

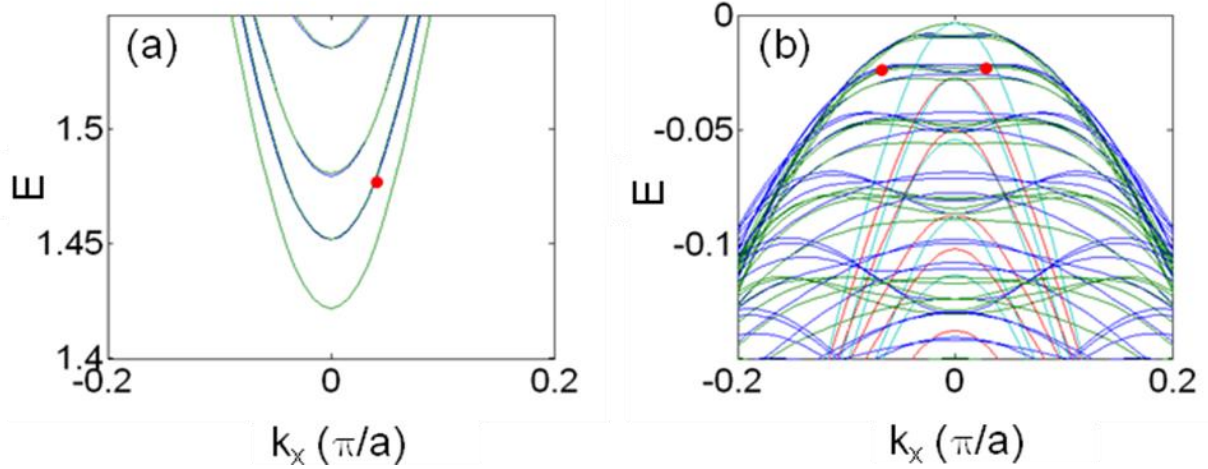

Figure 8. Energy dispersion relation showing the lowest (a) conduction bands and (b) valence bands of GaAs. ( y axis ranges from $E_C$ (or $E_V$) to $E_C$ (or $E_V$) + $5k_BT$ because $-\partial f_0/\partial E$ spread about $5k_BT$.) Each red dot represents a conducting channels for positive moving electrons at specific energy for an electron moving with a positive velocity. In the valence bands, most of the bands (especially heavy holes) have at least two conducting channels per energy

Table 1. Analytic "density-of-modes" and full band NEGF-TB simulation. For comparison, "density-of-states" effective masses ($m^*_{DOS}$) are also listed. The transport direction is along the $x$ direction. The electron ($m^e$) and hole effective masses ($m^{lh}$, $m^{hh}$) in the device coordinate ($x$, $y$, $z$) are used for analytic effective mass calculations and are given in units of the free electron mass. The 'heavy-hole' effective masses (Ge: 0.35 and GaAs: 0.51) assume spherical symmetry[75,76]. The extracted 'heavy-hole' effective mass for Ge and GaAs has a strong anisotropy (Ge: 0.17 [100], 0.37 [110], 0.53 [111], and GaAs: 0.38 [100], 0.66 [110], 0.84 [111]). CB denotes conduction band and VB denotes valence band. The top three shaded rows are for the conduction bands and generally show good agreement between analytic and numerically extracted values. The bottom three rows for the valence band (VB) generally show a much larger discrepancy. The two columns at the right (enclosed in dashed lines) show that analytic and numerically extracted density-of-states effective masses generally agree reasonably well, but the density-of-states effective masses are typically much lower than the density of modes effective masses.

| Material | $m^*_{DOM}$ | | $m^*_{DOS}$ | |
|---|---|---|---|---|
| | Analytic | Extracted | Analytic | Extracted |
| Ge CB | $4\sqrt{m^e_{yy} m^e_{zz}} = 1.18$ | 1.24 | 0.56 | 0.51 |
| GaAs CB | $m_{yy} = 0.066$ | 0.073 | 0.066 | 0.063 |
| Bi$_2$Te$_3$ CB | $2\sqrt{m^e_{xx} m^e_{zz}} + 4\sqrt{m^e_{yy} m^e_{zz}} = 1.18$ | 1.17 | 0.23 | 0.28 |
| Ge VB | $m^{lh}_{yy} + m^{hh}_{yy} = 0.37$ | 1.63 | 0.35 | 0.32 |
| GaAs VB | $m^{lh}_{yy} + m^{hh}_{yy} = 0.59$ | 0.97 | 0.52 | 0.39 |
| Bi$_2$Te$_3$ VB | $2\sqrt{m^h_{xx} m^h_{zz}} + 4\sqrt{m^h_{yy} m^h_{zz}} = 1.39$ | 3.53 | 0.36 | 0.41 |

Table 2. Summary of parameters used in Figure 6: fitted $\lambda_0 (nm)$ parameters, experimental lattice thermal conductivity $\kappa_l (Wm^{-1}K^{-1})$. In the power law form of the mean free path, $\langle\langle \lambda(E) \rangle\rangle = \lambda_0 \left(E/k_B T\right)^r$, $r$ is 0 since we assumed that phonon scattering is dominant.

| Material | $\lambda_0$ | $\kappa_l$ |
|---|---|---|
| Ge CB/VB | 29/9.5 | 58 |
| GaAs CB/VB | 110/39 | 55 |
| $Bi_2Te_3$ CB/VB | 18/4 | 1.5 |